# Analysis of video quality losses in the homogenous HEVC video transcoding

Tomasz Grajek, Jakub Stankowski, Damian Karwowski, Krzysztof Klimaszewski, Olgierd Stankiewicz, Krzysztof Wegner

Chair of Multimedia Telecommunications and Microelectronics

Poznań University of Technology

ul. Polanka 3, 60-965 Poznań

e-mail: {tgrajek, jstankowski, dkarwow, kklima, ostank, kwegner}@multimedia.edu.pl

*Abstract*—The paper presents quantitative analysis of the video quality losses in the homogenous HEVC video transcoder. With the use of HM15.0 reference software and a set of test video sequences, cascaded pixel domain video transcoder (CPDT) concept has been used to gather all the necessary data needed for the analysis. This experiment was done for wide range of source and target bitrates. The essential result of the work is extensive evaluation of CPDT, commonly used as a reference in works on effective video transcoding. Until now no such extensively performed study have been made available in the literature. Quality degradation between transcoded video and the video that would be result of direct compression of the original video at the same bitrate as the transcoded one have been reported. The dependency between quality degradation caused by transcoding and the bitrate changes of the transcoded data stream are clearly presented on graphs.

*Keywords—video compression; video transcoding; HEVC; H.265; CPDT; cascaded pixel domain video transcoder;*

*Highlights:*

- Deep analysis of video quality losses in the homogenous HEVC video transcoder.

- Presents evaluation of CPDT transcoder, as a reference for other works on video transcoding.

- Optimal conditions of video transcoding are pointed out.

- Theoretical analysis and explanation on optimal condition of video transcoding.

I. INTRODUCTION

Techniques of hybrid compression of digital video are the most commonly used method in which videos are transmitted in contemporary IT networks. This hybrid technique, that has been developed over the years, become a subject of numerous standardization projects of experts groups of ISO/IEC MPEG and ITU-T VCEG (e.g. MPEG-2 [1], H.263 [2], AVC [3, 4], HEVC [5, 6] standards). Constantly changing requirements imposed on video compression algorithms, as well as the improving capabilities of the hardware results in faster and faster changing generations of video compression technology.

Today, the state-of-the-art in the field of video compression is High Efficiency Video Coding (HEVC) technology [6], which has been jointly developed by ISO/IEC and ITU-T and published in 2013 simultaneously as an international standard of ISO/IEC MPEG-H part 2 and recommendation ITU-T H.265 [5]. The new technique is the successor of the widely used and extremely successful Advance Video Compression (AVC) technology [4, 7]. In relation to this technique, the HEVC allows up to 2-fold reduction in the size of the encoded images without compromising the quality of video [8], and, more importantly, it supports compression of ultra-high definition video, that is believed to be the direction in which the future video systems will evolve. For this reason, it is highly expected that within next few years, new HEVC technique will replace currently used AVC technology on near all fields.

Like the previous video compression standards (e.g. MPEG-2, AVC) the HEVC has been developed having many different applications in mind, i.e. not only digital television, but also mobile terminals, in which a codec has to run on devices with significantly different computing capabilities. For each of those applications and environments, the requirements differ, and, what follows, the set of compression tools used also varies. Very often, not only set of tools is limited but what more important the parameters of the encoded video itself are changed. For example, the target bitrate or resolution of the video must be suitable for the given application and target device. Thus, the need for the existence of different versions of the same video sequence with different parameters – resolution, bitrate, frame rate etc. Taking into account the fact, that after encoding of a sequence, the original video signal is no longer available (thus, there is no possibility to repeat the process of encoding from the original signal), the only way to produce a required bitstream with new parameters is to use a video transcoder. Because in the considered case of video transcoding only some of parameters of the encoded video are modified, and the general coding technology remains the same (in our case it is HEVC) this type of transcoding is called homogenous video transcoding. It is important to note here, that during transcoding a video, generally, it is only possible to obtain bitstream representing the video signal with lower parameters than the source bitstream and, in general, with lower video quality relative to the video signal decoded from the original bitstream. One of the goal of the research on the transcoding is to minimize the loss of quality caused by re-encoding (transcoding) the video in case when the previously lossy compressed video is used to create new bitstream at different target bitrate. Especially quality (losses) of such created video in comparison with the quality of a video encoded from the original sequence is interesting.

It is commonly known, that in the case of hybrid video compression, transcoding always introduces some video quality loss, regardless of the choice of the target bitrate of a transcoded video [9-13]. This is a direct result of lossy operations performed each time by a video encoder. Although research on the impact of video transcoding on the quality of video, as well as impact on the size of the encoded bitstream has already been carried out for the older compression techniques, (e.g. MPEG-2 or AVC) [9,12-16] no extensive study of a performance video transcoding have been presented in the literature so far. Moreover the conclusions of existing partial studies cannot be directly transposed to the new HEVC technology. The main reasons are the differences between the older and the new compression techniques that strongly affect the compression process. Some kind of exception here is the work [11] (published by some of the authors of this text), which shows results on accumulation of video quality losses in multiple HEVC encoding and decoding cycles. However, the above mentioned work also does not analyze in details the mechanism of occurring of video quality losses in process of HEVC transcoding. Among other things highlighted below the last issue is the subject of this work.

II. SPECIFIC OBJECTIVES OF THE WORK

Homogenous HEVC transcoding is the subject of research works for a few years now [17-24]. The main goal of the already published works was to reduce computational complexity of HEVC re-encoding by exploiting high similarity of selected parameters of the original (input) and the transcoded bitstreams. In these works, apart from the direct comparison of performance (computational complexity and coding efficiency) of the proposed solutions with the parameters of the reference pixel-domain cascaded transcoder the same phenomenon of video quality degradation in transcoder has not been the subject of detailed analysis. For this reason the authors of those works did not indicate what are the optimal conditions for HEVC video transcoding for which the loss of video quality is minimal.

The goal of this work is to perform deep study of the impact of video transcoding onto the quality of transcoded video material using the newest video compression standard. In this work the cascaded pixel domain HEVC video transcoder is considered. That is, the process of transcoding is based on feeding the once HEVC encoded and decoded video to another HEVC encoder (fig 1) and performing the full process of compression. As it was mentioned before, such a configuration is commonly used in literature as a reference in research on the improvement of video transcoding. Therefore, our work may be used as a reference for any future improvements of the transcoding process for HEVC data. The cascaded method can also be used in practice as the simplest possible way of transcoding.

During the work described in this article the authors try to answer two fundamental questions. First, what is the difference between the quality of a HEVC transcoded video (transcoded with a specific value of target bitrate) and the quality of a video that would be achieved after direct compression of the original video (assuming that such video could be made available) at the same target bitrate? What factors affect this change of quality? Secondly, if some optimal scenario of video transcoding exists (understood as a range of changes of bitrate in transcoder), for which the loss of video quality is minimal? Loss of quality again understood as the difference between the quality of a transcoded video (transcoded at some value of a target bitrate) and the quality of a video that would be achieved after direct compression of the original video at the same target bitrate. In order to better understand the mechanism of quality losses, short remarks on this topic are highlighted in the next section.

### III. Loss Of Quality in The HEVC Encoder – Short Remarks

Detailed, quantitative analysis of sources of video quality losses in the HEVC video encoder/transcoder is beyond the scope of this work. Therefore, only remarks clearing out the reasons of this loss are to be outlined in this section.

The degradation of quality during transcoding is owed to the fact, that HEVC is a lossy video coder, i.e. it omits some data in the process of encoding. There are three separate sources of those losses (quality degradation) in the HEVC encoder. The first one is Discrete Cosine Transform (DCT)-like transformation of residual data, that is not fully reversible. The second is quantization of the resulted transform coefficients. The third and the final source of losses is associated with modification of reconstructed image samples by in-loop deblocking and sample-adaptive offset (SAO) filters. They aim to improve the visual (subjective) quality, but, at the same time, they may introduce value shifts to the pixel values, and thus, decrease the objective quality of the image.

HEVC exploits finite precision approximation of DCT transformation [4]. Taking into consideration both – the coding efficiency and complexity aspects, HEVC standard requires 16 bit depth of values obtained at each stage of the transform process, including the sign bit (for 8-bit input image sample representation). In order to fulfill this requirement there is a necessity of bit shifting and clipping operations at some stages of residual data transformation. As a result, combination of forward (performed in the encoder) and inverse (calculated in the decoder) DCT-like transformations gives only near lossless reconstruction of input image samples, and not fully lossless. So, there is some (in practice very low) loss of quality. It should be noted however, that this quality loss happens twice in the transcoding procedure, when a sample goes through the DTC-like transformation chain: first during encoding of the original material and for the second time during transcoding itself.

The step that introduces the most losses in all hybrid video codecs is the quantization of transform coefficients of the residual data [6], HEVC not being an exception from this rule. The process of quantization is equivalent (however from mathematical point of view not exactly the same) to dividing a transform coefficients by quantization step size (so called *QStep*) together with subsequent rounding operations. This leads to irreversible loss of part of information about the signal – both the reconstructed and the original images may differ significantly from each other. The level of quality loss depends directly on quantization step size (on the value of *QStep*) – the higher the value, the coarser representation of transformed residual data. Coarse representation results in a lower quality of reconstructed images, but gives higher compression ratio of the video.

Due to the nature of the block based encoding used in the HEVC coder, the reconstructed image can exhibit some distortions, usually called as block artifacts, that are very annoying for the viewers, especially at lower bitrates. In order to improve the subjective quality of the reconstructed images, deblocking and SAO filters are used [6, 25, 26]. This helps to improve the subjective quality, but, at the same time, can degrade the objective quality metrics, such as the still commonly used PSNR.

In order to reduce the blocking artifacts (by the use of the deblocking filter) and attenuate the ringing artifacts (by the use of SAO filter) that are present in the reconstructed images after decoding, the values of image samples are appropriately modified before writing them to a file or showing them on a screen. These filtering operations, when carried out several times (which is the case in the transcoding scenario), may change the values of image samples essentially. Even though the goal of those filters is to improve the subjective quality, the effect of these operations on the PSNR metric (used to measure the quality in transcoding) is difficult to predict and the use of those filters may lead to the further reduction of objective quality metric value.

IV. LOSS OF QUALITY IN HOMOGENOUS HEVC TRANSCODING – METHODOLOGY OF EXPERIMENTS

Suppose that we have a HEVC encoded video material, compressed at *SourceBitrate*. We can transcode the bitstream by decoding it and re-encoding at a different *TargetBitrate*, as seen on Fig. 1. The question is: what is the difference between quality of a transcoded video *T* (obtained by re-encoding of a previously encoded video *R* to a new bitstream with rate equal to *TargetBitrate*) and the quality of a video *C* achieved by a direct encoding of the original video to a bitstream whose rate is also again equal to *TargetBitrate*? So, what actually loss of a video quality is introduced by using previously encoded video *R* instead of the original one *O*. What loss of quality is introduced by transcoding?

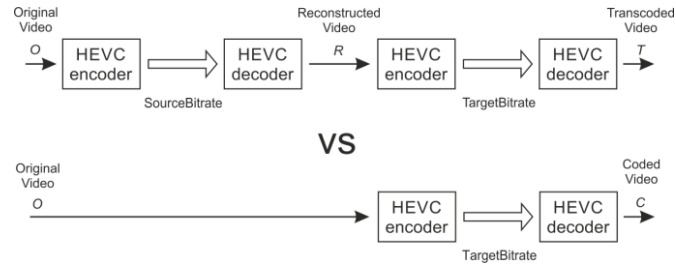

Fig. 1. Methodology of comparing quality losses during video transcoding .

In order to answer that question, an extensive and detailed experiment was conducted. At first, the original video materials $O$ have been encoded by HEVC encoder with the use of all of the possible quantization parameter values (QP ranging from 0 to 51) resulting in a number of bitstreams with different *SourceBitrate*. Resultant bitstreams have been decoded and the obtained video $R$ have been encoded again by HEVC reference encoder in order to simulate a cascade pixel domain transcoder. Re-encoding of a single decoded video $R$ has been done with the use of all possible quantization parameters (QP ranging from 0 to 51) in order to simulate a wide range of requested bitrate changes. As a result, more than 2500 video bitstreams have been prepared.

Quality of each decoded video, both after first encoding (video $R$) and after transcoding (video $T$), have been measured by luminance PSNR metric. Also, bitrate of each encoded stream has been gathered for further analysis.

The reference HEVC software has been configured according to the 'main_randomaccess' test conditions [27,28]. Such a scenario of video encoding is well recognized as high efficiency video compression scenario. Widely used *BlueSky*, *PedestrianArea*, *Riverbed*, *RushHour*, *Station2*, *Sunflower* full HD test sequences were taken as the test material (1920x1080 spatial resolution, 30 frames per second). In each case, only the first 200 frames from each sequence was encoded.

In order to determine the loss of image quality that is a result of cascaded video transcoding, ΔPSNR was calculated as the difference between the PSNR of transcoded video $T$ and the PSNR of encoded video $C$, both encoded at the same *TargetBitrate* (Fig. 1). The procedure for determining ΔPSNR is also illustrated in Fig. 2.

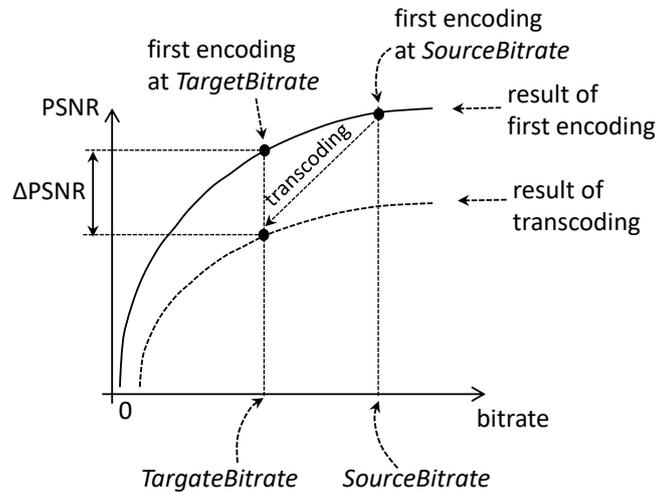

Fig. 2. The way of determining ΔPSNR in the homogenous HEVC transcoder.

In case of each transcoding, the relationship between the *SourceBitrate* (i.e. bitrate of the first video encoding) and the *TargetBitrate* (i.e. final bitrate of transcoded data) was additionally analyzed (see Fig. 2). Obtained partial results were used to determine ΔPSNR in the function of transcoding ratio (i.e. $\frac{TargetBitrate}{SourceBitrate}$ ratio) for each of the test sequences, and each value of source and target QP.

## V. Video Quality in Homogenous HEVC Transcoding – Results And Discussion

### A. Experimental results – general data

As evidenced by the results obtained during our experiments, transcoding of the HEVC bitstreams leads to a noticeable degradation of the quality of video. Fig. 3 presents exemplary but representative data of video transcoding that were obtained for *BlueSky* test sequence. In this figure, the solid line represents 'PSNR – bitrate' curve which is a result of a first step of the experiment: direct encoding of original test sequence with various QPs. A few selected values of QP = 22, 28, 32, 38 have been directly marked on the plot using rectangles. This line represents quality of a video *C* (from Fig. 1) encoded at different *TargetBitrate*.

The dotted lines give information about the results of transcoding of source bitstream to new bitstreams with different rate. Each dotted line represents a PSNR-bitrate curve of a transcoded video produced based on a single selected source bitstream encoded at some *SourceBitrate*. Those source bitstreams were the bitstreams obtained in the first step of encoding for the QP values of 22, 28, 32, 38.

The reasons why the quality of the transcoded video is always worse than the quality of a source video have already been outlined in Section III.

It must be emphasized here, that although the overall nature of changes of quality of video follows the trends seen for the exemplary sequence (see Fig. 3), the exact results of transcoding also depends on the content of a video to some extent.

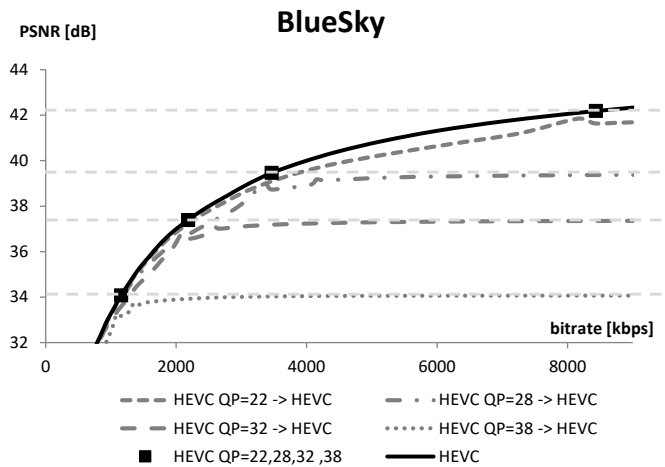

Fig. 3. Exemplary results of video transcoding obtained for *BlueSky* test sequence.

The general relationship of quality differences ΔPSNR (quality loss) in a function of transcoding ratio - $\frac{TargetBitrate}{SourceBitrate}$ is presented in Fig. 4. In this figure, each dot represent the result of a single transcoding, from a bitstream encoded at *SourceBitrate* to a bitstream encoded at *TargetBitrate*. The solid line is a result of averaging of the results that are marked with dots. As can be easily seen, the value of $\frac{TargetBitrate}{SourceBitrate}$ ratio affects the results to a high extent.

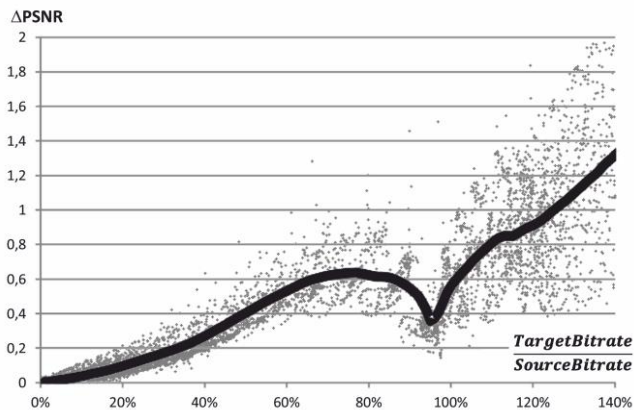

Fig. 4. ΔPSNR (quality loss) with respect to transcoding ratio (i.e. *TargetBitrate*/*SourceBitrate* ratio). The result of averaging the partial data obtained for individual test sequences and different values of *SourceBitrate*.

The main conclusion that can be drawn here is that the lower the value of transcoding ratio (which translates into bigger difference between *TargetBitrate* and *SourceBitrate* values), the lower the loss of quality expressed by ΔPSNR, as compared to a direct encoding of original data at *TargetBitrate*.

In the extreme case, in which the *TargetBitrate* is very small while the *SourceBitrate* is significantly larger (so, value of the ratio is small as well), transcoding will not lead to a significant degradation of a video quality (understood as ΔPSNR). This can be explained by the fact that encoding video material at very small *TargetBitrate* causes big quality losses. In such a condition, it is not relevant whether we use original video material or previously slightly degraded by encoding at relatively big *SourceBitrate*.

In the opposite case of high values of *TargetBitrate* (which also means higher values of the transcoding ratio), the quality of the used video material (original or previously encoded at similar to *TargetBitrate, SourceBitrate*) plays significant difference. This highlights greater loss of video quality during transcoding with high transcoding ratio, when compared to the scenario with small transcoding ratio.

Summarizing, we can say that an average maximum PSNR decrease of 1.4dB was observed (in comparison to video encoded directly from the original material, at the same *TargetBitrate*) for the range of the transcoding ratios considered in the research. One has to remember, though, that this average maximum value also includes results for cases where the target QP was smaller than the source QP, so the *TargetBitrate* was larger than the *SourceBitrate* – a scenario which rarely finds any practical justification. In cases which have the greatest practical use (that is $\frac{TargetBitrate}{SourceBitrate} < 100\%$) an average maximum decrease of PSNR is below 0.7dB.

Another observation concerns the results that were obtained in cases of a moderate reduction of bitrate in transcoder. What is very interesting, there is an optimal point of transcoding, i.e., optimal value of transcoding ratio for which the loss of video quality finds its local minimum in this transcoding scenario. The averaged results (see Fig. 4) show that this point is $\frac{TargetBitrate}{SourceBitrate} = 95\%$ and the loss of video quality is in the range of 0.35 dB for this point. Deeper analysis of this phenomena will be a topic of subsection C of this section.

Besides the point that "minimizes" the loss of video quality (for a moderate reductions of bitrate), there is also point of transcoding which forms a local maximum of the loss of video quality. This is for the case of the transcoding ratio equal approximately to 75%, which gives the loss of video quality ΔPSNR = 0.63 dB. Consequently, reduction of the bitrate by 25% (more or less) results in the highest loss of video quality, when compared to the reference direct encoding results.

Another obvious observation that arises from the analysis of the graph is that the ratios above 100%, for which the *TargetBitrate* is larger than the *SourceBitrate*, are, in average, characterized by the steadily increasing quality loss. The larger the ratio the higher the loss. The explanation here is that for a larger target bitrate, the transcoding is not able to compete with the direct encoding, since the direct encoding of the original data is able to preserve the details that are simply not present in the encoded data (video *R* from Fig. 1) fed to the transcoding chain. Therefore, being unable to recover those details, the transcoder is not able to produce the video of the similar quality as in the reference scenario.

B. *Experimental results – more detailed data*

The exact relation between expected ΔPSNR and the transcoding ratio depends on the content of a sequence to some extent, but, in general, the overall properties of data remain unchanged. The results that were obtained for individual test sequences are presented in Fig. 5.

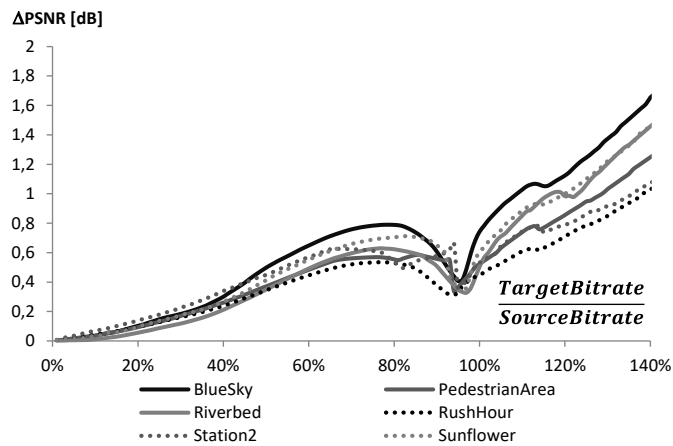

Fig. 5. The results of averaging the partial data obtained for the considered values of QP. Results of individual test sequences.

Deeper analysis of the presented relationships allowed to conclude that higher losses of quality occur in case of a more complex (in terms of image content) video sequences. In this case it is more difficult to predict the content of image blocks which results in higher values of prediction error of image samples that undergo quantization. In the described scenario, the quantization of a prediction error signal causes a larger loss of video quality compared to the sequences with a simple content. In the latter case, smaller prediction error signal is obtained, and that results in a smaller losses of information observed in the process of data quantization. It is also worth noting that for complex sequences, even small QP (high *SourceBitrate*) used for the first compression in the process, removes significant amount of data, which are impossible to restore before the second encoding in the cascade transcoder.

The relation between expected quality loss ΔPSNR and the transcoding ratio depends even stronger on the value of *SourceBitrate* (i.e. on the value of QP used in first stage of cascade transcoder) than on the content of a video itself. As can be seen from exemplary results presented for the *BlueSky* sequence (see Fig. 6), the lower the *SourceBitrate* (the lowest bitrate for encoding with QP=38 – round dots, the highest for QP = 22 – short dense dashes), the higher the expected losses of video quality when transcoding a video. This is particularly visible for higher values of the transcoding ratio. More detailed analysis of this property revealed that reasons for this can be found in the nature of the PSNR-bitrate curve (see solid line in Fig. 3). This curve is very steep for low values of bitrates (which corresponds to bigger values of QP), and relatively flat for high bitrates (so, for low QP values). As a result, transcoding of low bitrate bitstreams will result in a greater change (loss) of video quality, than in the case of higher values of bitrates. Complementary to the results of Fig. 6 are data presented in Fig. 7 and 8, presenting the results obtained for individual test sequences for the two strongly different values of QP: 28 and 38. The presented results confirm all the conclusions that were formulated before.

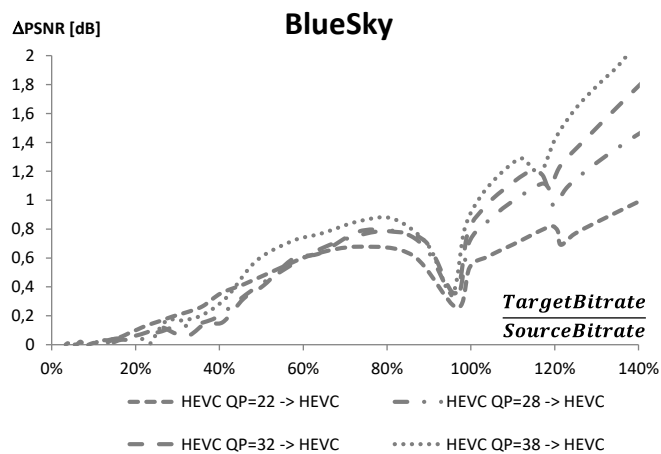

Fig. 6. ΔPSNR with respect to transcoding ratio for four scenarios of encoding the original video: QP=22, 28, 32, 38.

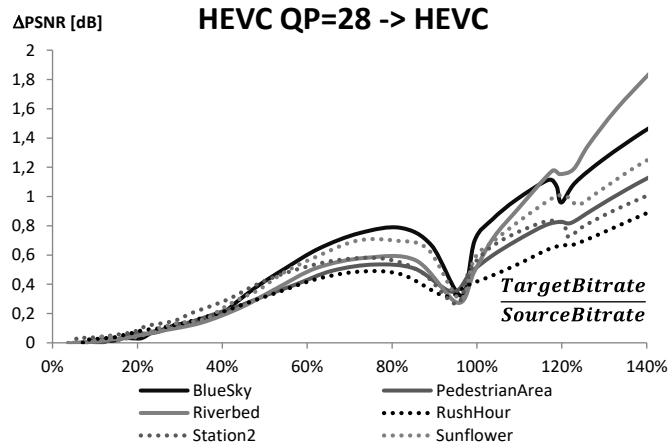

Fig. 7. ΔPSNR with respect to transcoding ratio for individual sequences. Results for first step compression with QP=28.

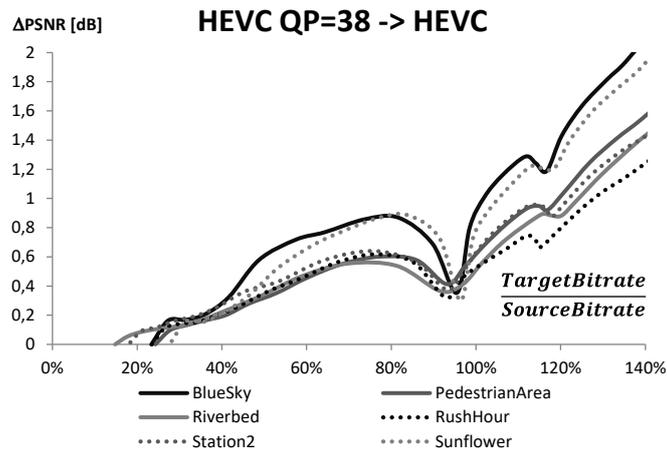

Fig. 8. ΔPSNR with respect to transcoding ratio for individual sequences. Results for first step compression with QP=38.

*C. Conditions for minimum loss video transcoding*

As it was pointed out in the subsection A, there exists a characteristic point of video transcoding which minimizes the loss of video quality for scenario of a moderate reduction of bitrate in a video transcoder. According to the averaged experimental data, this point is $\frac{TargetBitrate}{SourceBitrate} = 95\%$. In order to investigate causes of the existence of such a point, each of the partial data has been carefully analyzed in terms of settings of the encoder which led to individual results. In particular, relationship between value of quantization parameter $QP_S$ used for encoding of the original video (that produces video *R* from Fig. 1) and value of this parameter $QP_T$ used later in the transcoder were studied. Results of this analysis achieved for range of the transcoding ratio from 75% to 100% were presented in Fig. 9.

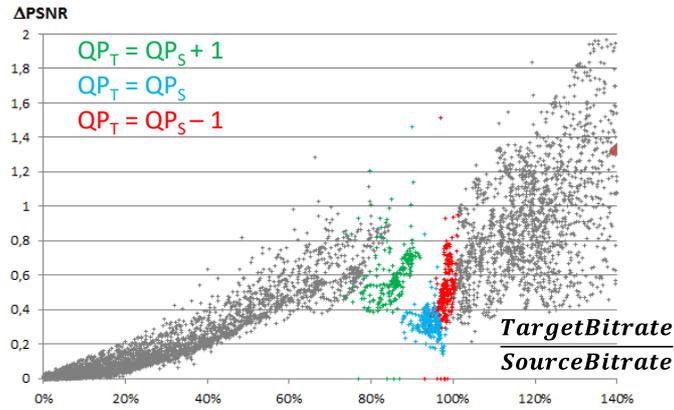

Fig. 9. ΔPSNR with respect to transcoding ratio (i.e. *TargetBitrate/SourceBitrate* ratio). Figure presents partial data obtained for individual test sequences.

As it turned out, the local minimum in the quality loss is obtained for the case when the encoding in the second step of transcoding uses the same QP value as the one used in the first step of encoding. The points closest to this minimum correspond to the cases when the second step encoding was performed with the QP value higher or lower by 1 than the QP used in the first step. Partial results obtained for the various scenarios (mutual relationship of values of $QP_T$ and $QP_S$) have been marked with appropriate distinctive color in Fig. 9.

The important thing to note is that the bitrate ratio of 100% is obtained almost exclusively for the cases when the encoding in the second step is performed with the QP smaller by 1 than the value used in the first step, while causing a loss of 0.5dB in quality on average.

## VI. Video Quality in Homogenous HEVC Transcoding – Further Considerations

Presented in the previous section results clearly show that when transcoding a video the mutual relationship of the quantization parameters of a source bitstream $QP_S$ and target bitstream $QP_T$ determines directly how big the loss of quality of a signal will be. Fig. 9. presents the empirical data of video quality loss for cases when $QP_T = QP_S$ and $QP_T = QP_S \pm 1$ in homogenous HEVC transcoder. But at this point there are still open questions. First, what is the justification of the nature of obtained results? And secondly, how big the loss of video quality would be when the mutual relation of values of $QP_S$ and $QP_T$ would be different to those already analyzed in Section C of the previous point? These issues will be the subject of a more detailed discussion in the following paragraphs.

*A. The nature of obtained results – analysis of reasons*

In the context of the first question, a theoretical analysis has been done to explore the impact of data quantization on the size of quantization error. This analysis has been done assuming the special case of two consecutive stages of quantization. In the considered scenario, the first stage of data quantization is realized with a parameter $QP_S$, while the second stage with a parameter $QP_T$. The following special cases were considered in the analysis: 1) $QP_S = QP_T$, 2) $QP_S < QP_T$, and finally 3) $QP_S > QP_T$.

Conclusions from the analysis have been presented on illustrations of Fig. 10, from which the following general conclusions can be drawn.

The use of the two identical quantizers in both the stages (thus, $QP_S = QP_T$) will lead to a value of final quantization error 'e' whose value will never be greater than the maximum error '$e_{max}$' for the second quantizer (see Case A in the Fig. 10). This is because of equal quantization step sizes used in both the first and the second quantizer. Of course, this will be the case when the output value of the first quantizer will be directly put onto the second quantizer (such a scenario was adopted in Fig. 10). However, in a real transcoder the situation is somewhat different. The data that are quantized in a transcoder (with quantization parameter value $QP_T$) may be different from those quantized in the first stage of video encoding (that is realized with the parameter $QP_S$), due to the other steps of processing, like nonlinear filtering and prediction that will influence the final value of the sample. This change of value explains the non-zero loss of video quality for $QP_S = QP_T$ scenario (as presented in Fig. 9). But the general theoretical conclusion highlighted above that the full compliance of the two quantization step sizes ($QP_S = QP_T$) gives the minimum error of video transcoding is still true, compared with other relation of values of $QP_S$ and $QP_T$.

In the cases when the two quantizers are different from each other (i.e. $QP_S$ differs from $QP_T$) the final quantization error 'e' can be bigger than the maximum error '$e_{max}$' of the second quantizer. The discussed situation for the two separate cases in which $QP_S < QP_T$ or $QP_S > QP_T$ has been illustrated with Cases B and C in Fig. 10. The presented conclusion results from the fact of different allocation of decision boundaries (on axis of quantized values) in the first and the second quantizer, which follows directly from different width of quantization intervals. Thus, in these cases the resulting error of transcoding is generally greater than in the case when $QP_S = QP_T$.

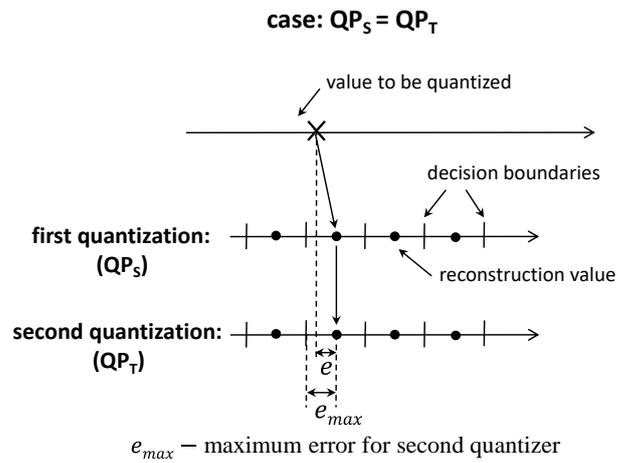

Case A - $QP_S = QP_T$

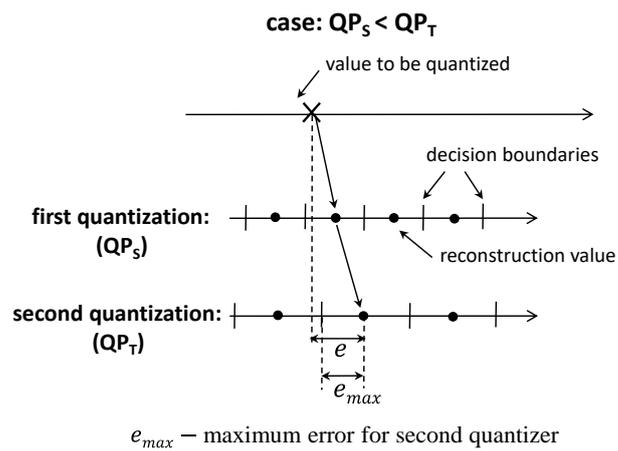

Case B - $QP_S < QP_T$

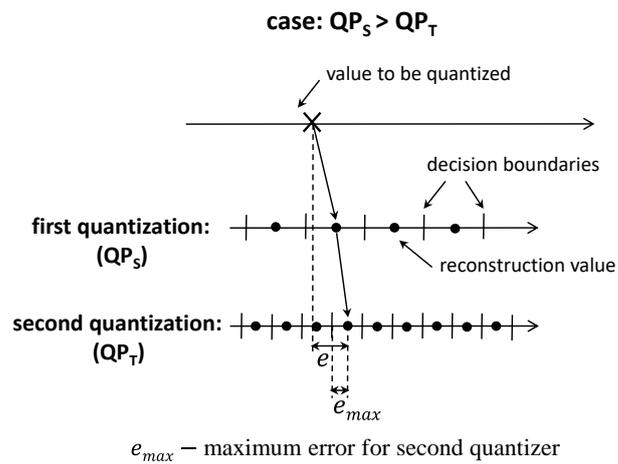

Case C - $QP_S > QP_T$

Fig. 10. Impact of the relation of values of $QP_S$ and $QP_T$ on the size of quantization error. Analysis of three cases: Case A - $QP_S = QP_T$, Case B - $QP_S < QP_T$, Case C - $QP_S > QP_T$.

## B. Relation of $QP_T$ to $QP_S$ and its impact on errors of transcoding

The illustrations from the previous point clearly present the mechanism of introducing the transcoding errors in a video transcoder when QPs differs from $QP_T$, but there is still unsolved question about the magnitude of this error in practical situations. In order to answer the question of error magnitude for certain cases, the range of values of QPs and $QP_T$ were tested. This way the influence of the width difference of the quantization step at different stages of encoding was tested in an additional experiment.

In this experiment the uniform quantizer with quantization step equal to $QStep_S$ was used to quantize every possible transform coefficient value from a HEVC encoder. Then, the dequantized value was quantized once again using quantization step equal to $QStep_T$. For the reference purposes, the same initial value of transform coefficient was directly quantized using the quantization step equal to $QStep_T$. For both cases, the quantization error $E$ was calculated – $E_a$ and $E_b$ respectively for those two cases. The example calculation is shown on Fig. 11.

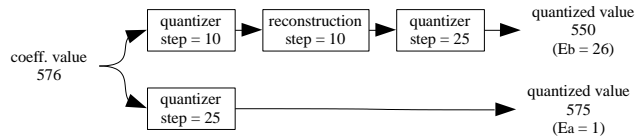

Fig. 11. An example result of the experiment with varying quantization step.

The average value of errors $E_a$ and $E_b$ for the example case ($QStep_S$=10 and $QStep_T$=20) (averaged over all possible coefficient values) are 12 and 14.5 respectively. Therefore it can be seen that transcoding introduces additional error when compared to a case with direct compression with a given width of the quantization step, which agrees with the observations described in the previous chapters. The ratio of mean error values (that is $E_b/E_a$) is 1.2.

To examine the influence of quantized step width change on the results, a different experiment was conducted, for which the $QStep_S$ was kept constant at value of 12, while the $QStep_T$ was changed from 2 to 40 with a step of 1. The ratio of mean error value $E_b / E_a$ is given as a function of $QStep_T$ on Fig. 12.

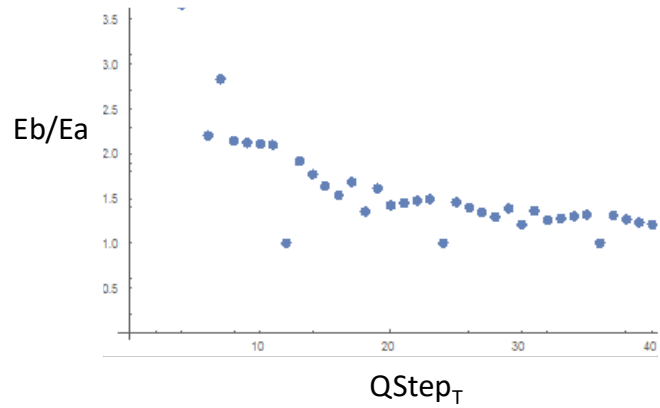

Fig. 12. The error ratio as a function of $QStep_T$.

What is evident from that figure, is the following. The minimum error can be obtained only when the $QStep_T$ width is an integer multiply of the $QStep_S$ width. For such cases the error remains the same in both scenarios, therefore $E_b / E_a = 1$. For every other case, the scenario with requantization gives higher average error, even when the $QStep_T$ changes by 1 from the multiply of $QStep_S$. For such a case, the error can be more than twice that high.

Another observation also agrees with previously presented results – the higher the $QStep_T$, the lower the $E_b / E_a$ ratio, therefore the requantization loss is smaller.

On the Fig. 13, the relationship between the $E_b / E_a$ coefficient and both quantization steps, $QStep_S$ and $QStep_T$, is visualized.

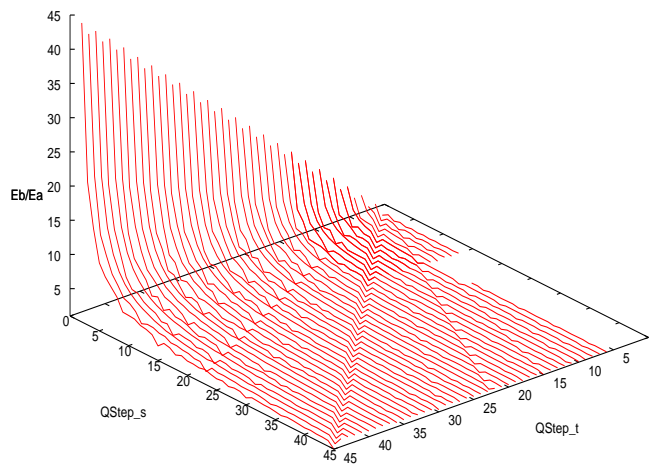

Fig. 13. The error ratio as a function of $QStep_T$ and $QStep_S$.

The minimum values for $QStep_S = QStep_T$ are clearly noticeable on this figure, as well as less noticeable minimum for $QStep_T / QStep_S = 1.5, 2.5, 3.5$ and so on.

On the Fig. 14, the same data are plotted in 2D – again, the minimum mentioned above are clearly visible.

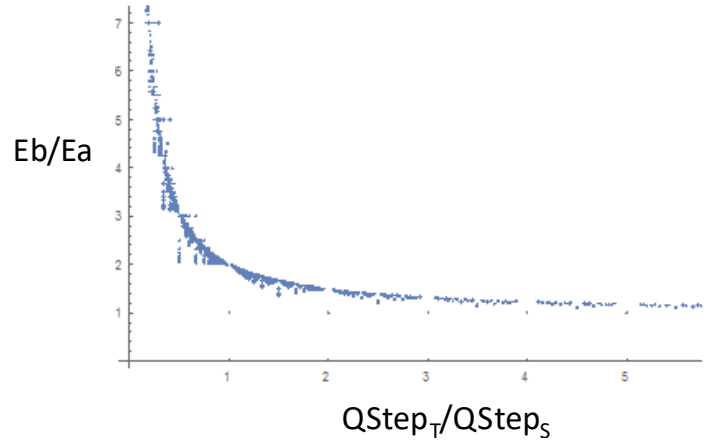

Fig. 14. The error ratio as a function of $QStep_T$ / $QStep_S$ ratio.

The reason for the minimum to occur for $QStep_T$ being a multiple of $QStep_S$ is obvious: the quantization step limits (decision boundaries) are overlapping for those cases, therefore the samples are always represented by the same quantized value, regardless of whether they underwent a single quantization with $QStep_T$ or two consecutive quantizations – first with $QStep_S$ and then with $QStep_T$. This is presented on Fig. 15.

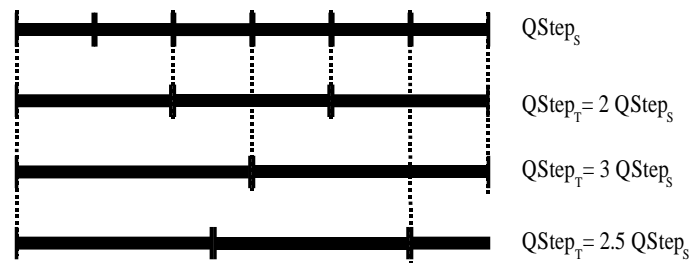

Fig. 15. The quantization step limits overlap for different $QStep_T$ / $QStep_S$ ratios.

The reason for the $E_b$ / $E_a$ ratio for $QStep_T = 2.5\ QStep_S$ not being 1 is that only every second of the decision boundaries overlap. This means that 2 out of 3 steps will overlap entirely, but every third one will get requantized and will be assigned to a certain step of $QStep_T$, with maximal quantization error equal to 0.5 $QStep_S$, as it is shown on Fig. 16. The dark gray area marks the range of input values that are assigned to a different step in case of requantization.

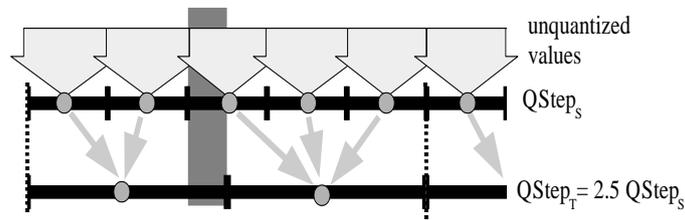

Fig. 16. The requantization introduced additional error – marked in dark gray.

## VII. FINAL REMARKS AND CONCLUSIONS

A quantitative analysis of expected quality losses in the homogenous HEVC transcoder was presented in the paper. Presented results can be used as a reference for the future research on video transcoder improvements. The results have shown that, although transcoding of video always leads to some loss of a video quality, there are transcoding conditions for which the loss of quality is minimal - with respect to the quality of video encoded from original video at same bitrate as the transcoded video. The minimum quality loses of ΔPSNR=0.35 dB is realized with a moderate reduction of bitrate, it is for transcoding ratio $\frac{TargetBitrate}{SourceBitrate} = 95\%$. It corresponds to the case of using the same value of QP in first and second video encodings. It should be further noted that exact relation ΔPSNR versus transcoding ratio depends significantly on *SourceBitrate*, as well as the content of a video sequence.

## ACKNOWLEDGMENT

Research project was supported by The National Centre for Research and Development, POLAND, Grant no. LIDER/023/541/L-4/12/NCBR/2013.